\documentclass[12pt]{article}
\usepackage{amsmath}
\usepackage{graphicx}
\usepackage{enumerate}
\usepackage{natbib}
\usepackage{url} 
\usepackage{pdflscape}
\usepackage{tikz}
\usetikzlibrary{positioning,fit}

\usepackage{amsmath}
\usepackage{graphicx,psfrag,epsf}
\usepackage{enumerate}
\usepackage{natbib}
\usepackage{url} 
\usepackage{amsthm}
\usepackage{booktabs}
\usepackage{framed}  
\usepackage{caption}
\usepackage{pgfplots}
\usepgfplotslibrary{dateplot}
\usetikzlibrary{snakes}
\usepackage{float}
\usepackage{amsfonts}
\usepackage{multirow, booktabs}
\usepackage{cleveref}
\usepackage{subcaption}
\usepackage[ruled,vlined]{algorithm2e}
\usepackage[T1]{fontenc}
\usepackage[utf8]{inputenc}
\usepackage{authblk}
\usepackage[multiple]{footmisc}
\usepackage{blindtext,titlefoot}
\usepackage{sectsty}
\usepackage{xcolor}
\usepackage{tikz}
\usepackage{amsmath}
\usepackage{graphicx}
\usepackage{comment}
\usepackage{amsfonts}
\usepackage{bbm}
\usepackage{amssymb}
\usepackage{setspace}
\usepackage{natbib}
\usepackage[figuresright]{rotating}
\usepackage{adjustbox}
\usepackage{pifont}
\usetikzlibrary{positioning, shapes.geometric}

\usepackage{tabularx}
\usepackage{ragged2e} 
\newcolumntype{L}{>{\RaggedRight}X} 
\usepackage{lipsum} 

\usepackage[ruled,vlined]{algorithm2e}
\usepackage{amsmath, amssymb}
\usepackage{amsfonts, multirow, epsfig, subfig}
\usepackage{graphicx, pdflscape, verbatim, enumerate, colortbl, setspace}
\usepackage{setspace, color,bm}
\usepackage[normalem]{ulem}
\usepackage{cite}
\usepackage{multirow}
\usepackage{booktabs,array}
\usepackage{url}
\usepackage{bbm}

\newtheorem{proposition}{Proposition}
\theoremstyle{definition}
\newtheorem{condition}{Condition}

\newtheorem{theorem}{Theorem}

\theoremstyle{definition}
\newtheorem{remark}{Remark}

\makeatletter
\newcommand*{\rom}[1]{\expandafter\@slowromancap\romannumeral #1@}
\makeatother

\begin{document}

\def\spacingset#1{\renewcommand{\baselinestretch}%
{#1}\small\normalsize} \spacingset{1}

\sectionfont{\bfseries\large\sffamily}%
%
\newcommand*\emptycirc[1][1ex]{\tikz\draw (0,0) circle (#1);} 
\newcommand*\halfcirc[1][1ex]{%
  \begin{tikzpicture}
  \draw[fill] (0,0)-- (90:#1) arc (90:270:#1) -- cycle ;
  \draw (0,0) circle (#1);
  \end{tikzpicture}}
\newcommand*\fullcirc[1][1ex]{\tikz\fill (0,0) circle (#1);} 

\subsectionfont{\bfseries\sffamily\normalsize}%
%


\def\spacingset#1{\renewcommand{\baselinestretch}%
{#1}\small\normalsize} \spacingset{1}

\begin{center}
    \Large \bf A Universal Framework for Factorial Matched Observational Studies with General Treatment Types: Design, Analysis, and Applications
\end{center}

\let\thefootnote\relax\footnotetext{$^{*}$Address for Correspondence: Siyu Heng, Department of Biostatistics, School of Global Public Health, New York University, New York, NY 10003, U.S.A. (Email: siyuheng@nyu.edu).}

\let\thefootnote\relax\footnotetext{This version of the manuscript focuses on presenting the statistical methodology and a preliminary protocol for the data analysis. The outcome analysis results will be incorporated in a subsequent version. }

\begin{center}
  \large $\text{Jianan Zhu}^{1}$, $\text{Tianruo Zhang}^{2}$, $\text{Diana Silver}^{3}$, $\text{Ellicott C Matthay}^{4}$, $\text{Omar El-Shahawy}^{4,5}$, $\text{Hyunseung Kang}^{5}$, and $\text{Siyu Heng}^{*, 1}$
\end{center}

\begin{center}
   \large \textit{$^{1}$Department of Biostatistics, School of Global Public Health, \\ New York University}
\end{center}

\begin{center}
   \large \textit{$^{2}$Technology \& Operations Management Unit,
    Harvard Business School,
    Harvard University}
\end{center}

\begin{center}
   \large \textit{$^{3}$Department of Public Health Policy and Management, School of Global Public Health, New York University}
\end{center}

\begin{center}
   \large \textit{$^{4}$Department of Population Health, Grossman School of Medicine,\\ New York University }
\end{center}

\begin{center}
   \large \textit{$^{5}$Department of Global and Environmental Health, School of Global Public Health, New York University}
\end{center}

\begin{center}
   \large \textit{$^{6}$Department of Statistics, School of Computer, Data \& Information Sciences, University of Wisconsin-Madison }
\end{center}

\begin{abstract}
Matching is one of the most widely used causal inference frameworks in observational studies. However, all the existing matching-based causal inference methods are designed for either a single treatment with general treatment types (e.g., binary, ordinal, or continuous) or factorial (multiple) treatments with binary treatments only. To our knowledge, no existing matching-based causal methods can handle factorial treatments with general treatment types. This critical gap substantially hinders the applicability of matching in many real-world problems, in which there are often multiple, potentially non-binary (e.g., continuous) treatment components. To address this critical gap, this work develops a universal framework for the design and analysis of factorial matched observational studies with general treatment types (e.g., binary, ordinal, or continuous). We first propose a two-stage non-bipartite matching algorithm that constructs matched sets of units with similar covariates but distinct combinations of treatment doses, thereby enabling valid estimation of both main and interaction effects. We then introduce a new class of generalized factorial Neyman-type estimands that provide model-free, finite-population-valid definitions of marginal and interaction causal effects under factorial treatments with general treatment types. Randomization-based Fisher-type and Neyman-type inference procedures are developed, including unbiased estimators, asymptotically valid variance estimators, and variance adjustments incorporating covariate information for improved efficiency. Finally, we illustrate the proposed framework through a county-level application that evaluates the causal impacts of work- and non-work-trip reductions (social distancing practices) on COVID-19-related and drug-related outcomes during the COVID-19 pandemic in the United States. 
\end{abstract}

\spacingset{1.73} 

\section{Introduction}

\subsection{Background}

In causal inference, randomized experiments are the gold standard. However, due to ethical or budgetary concerns, conducting a randomized experiment is not practical in many settings, and an observational study is a natural alternative. Among various observational study designs, matching is among the most widely used ones. A matched observational study seeks to mimic randomized experiments using observational (non-experimental) data by forming matched sets of units with similar covariate values but different treatment values (e.g., treated versus control in the binary treatment case or higher versus lower treatment doses in the continuous treatment case). After matching, the treatments are as-if randomly assigned within each matched set, enabling valid randomization-based causal inference as in a randomized experiment \citep{rosenbaum2002observational,rosenbaum2020design,rubin1973matching,rubin1974estimating,rubin2007matching,imbens2015causal,ding2024first,stuart2010matching}. Through a sequence of seminal works, the design and analysis methods for matched observational studies have been well-established for datasets with a single or factorial (multiple) binary treatments \citep{rubin1979,rosenbaum2002observational,lopez2017multiple,dasgupta2015factorial,nattino2021triplet}, as well as for datasets with a single non-binary (e.g., continuous) treatment \citep{lu2001matching,greevy2023nonbipartite,zhang2023matching,zhang2024bridging}. However, in many real-world applications, observational data often involve factorial non-binary (e.g., continuous) treatment variables that can potentially influence the outcome. In such cases, researchers are usually interested in studying the causal effects (e.g., main and interaction effects) of multiple treatment components (potentially non-binary or continuous) simultaneously. However, there is no established study design and statistical analysis approach for factorial matched observational study with general treatment component types (e.g., binary, ordinal, or continuous). Because of this gap, when handling, for example, multiple non-binary continuous treatments, previous matched observational studies had to either dichotomize the continuous treatments \citep{fong2018continuous, Boyd2010sex, Nielsen2011aid,fogarty2019biased,yu2023matching} or combine multiple continuous treatments into a single, compound continuous treatment using some ad hoc weights \citep{zhang2023social,frazier2024bias,pan2020sdi,zhang2021platform}.

\subsubsection{Motivating Example}

During the COVID-19 pandemic, social distancing was widely used as one of the most effective strategies to mitigate disease transmission. To comply with enacted federal policies, a variety of interventions were implemented across different societal levels to promote physical distancing and reduce interpersonal contact. For example, companies such as Google, Twitter, and Facebook transitioned to remote work (i.e., reductions in work trips), while public restaurants and transportation systems reduced seating capacity to support social distancing guidelines (i.e., reductions in non-work trips). As a result, multiple components of human mobility should be considered jointly when evaluating the extent to which social distancing was practiced. Reductions in work and non-work trips represent two key mobility metrics used in such evaluations.

Given the widespread implementation of social distancing policies, a growing body of research has sought to examine the causal relationship between social distancing behaviors and health-related outcomes using observational data. Among those observational studies, matching has been one of the most widely used frameworks for causal inference (\citealp{zhang2023social,frazier2024bias}). In practice, because of a lack of valid matching designs for handling multiple continuous treatments (e.g., reductions in work and non-work trips), previous observational studies often used some prespecified weights to combine multiple continuous treatments into a single composite continuous measure to facilitate matching and subsequent randomization-based (design-based) causal inference (\citealp{zhang2023social,frazier2024bias,pan2020sdi,zhang2021platform}). However, this common practice can be problematic since these weights are often based on background knowledge and lack statistical justification, and may not have a well-defined causal interpretation. For example, if the potential outcomes under originally factorial (multivariate) treatments cannot be reduced to a potential outcome function in terms of a one-dimensional composite exposure, such a composite treatment measure does not correspond to any well-defined potential outcomes (see Section~\ref{sec: review} for details). Moreover, because such practice collapses the multi-dimensional treatments into a one-dimensional composite measure, researchers were \textit{unable} to study the interaction effects of different treatments.

\subsection{Our Contributions}\label{sec: contribution}

Motivated by the aforementioned lack of valid and flexible matching designs and associated causal inference methods for factorial observational studies with general treatment types, as well as the practical consequences of such methodology gaps in the existing COVID-19 social distancing policy evaluation practices, in this work, we propose a novel and universal design and inference framework for factorial matched observational studies (i.e., matched observational studies involving multiple treatments) with general treatment types (e.g., binary, ordinal, or continuous). Specifically, our contributions cover the following aspects: 
\begin{itemize}

    \item \textbf{(Study Design)} In the design stage, we propose an innovative universal matching procedure for observational studies with multiple general treatment variables. We also introduce new, finite-population-valid Neyman-type estimands to capture both marginal and interaction effects in the study.

    \begin{itemize}
        \item \textbf{(Matching Procedure)} First, we introduce a multi-stage non-bipartite matching procedure for factorial observational studies with general treatment types. Unlike existing approaches that rely on dichotomization or aggregation to facilitate matching under non-binary (e.g., ordinal or continuous) treatments, our method preserves the original treatment scales by directly matching units based on their treatment doses. This design lays the foundation for valid and interpretable estimation and inference of the causal effects (main and interaction effects) associated with all the treatment components.
    
    \item \textbf{(Causal Estimands)} Second, recognizing the presence of multiple treatment variables, we consider not only the marginal causal effects of each treatment but also their interaction effects. To formally capture these effects, we propose a new class of estimands, called generalized factorial Neyman-type estimands, for factorial matched observational studies with general treatment types. To our knowledge, these estimands are the first model-free and finite-population-valid causal estimands that can accommodate settings with multiple, possibly non-binary treatments in factorial observational studies.
    \end{itemize}


     \item \textbf{(Causal Analysis)} Next, we propose novel randomization-based (design-based) inference frameworks for factorial matched observational studies with general treatments, covering both Fisher-type and Neyman-type inference modes. 
     \begin{itemize}
         \item \textbf{(Hypothesis Testing)} First, we propose Fisher-type randomization tests for testing the dose-response relationship between the factorial treatment variables and the potential outcomes, including Fisher's sharp null hypothesis of no effect as a special case. The proposed randomization tests are finite-population-exact and can accommodate a wide range of test statistics.
         
         \item \textbf{(Estimation)} Second, we introduce randomization-based Neyman-type estimators for the proposed generalized factorial Neyman-type estimands. These estimators are all unbiased for estimating the average main and interaction effects of multiple treatments, conditional on matching and the ignorability assumption.
         
         \item \textbf{(Inference)} Third, we derive asymptotically valid variance estimators for all proposed randomization-based Neyman-type estimators in the context of matched observational data. We show that the proposed variance estimators can facilitate finite-population asymptotically valid inference for the proposed generalized factorial Neyman-type estimands. We also discuss how to incorporate the covariate information into the inference procedure to further improve efficiency while maintaining statistical validity. 
         
     \end{itemize}

    \item \textbf{(Application)} Finally, we will apply the proposed design and inference frameworks to explore the county-level causal effects of social distancing practices on public health outcomes. To support this investigation, we integrate data from the University of Maryland's COVID-19 Impact Analysis Platform, the United States Census Bureau, the County Health Rankings and Roadmap Program, and the Centers for Disease Control and Prevention's (CDC's) county-level mortality data. Among various measures of social distancing, we focus on two distinct continuous treatment doses: the county-level average per-person percentage reductions in work and non-work trips during the pandemic, compared to the pre-pandemic baseline. The primary outcome is the cumulative number of COVID-19 cases during the pandemic. The secondary outcomes include the cumulative number of COVID-19-related deaths and overdose-related deaths over the same period.

\end{itemize}


\section{A Brief Review of Existing Methods and Their Limitations}\label{sec: review}

Suppose there is an $M$-dimensional factorial treatment (i.e., $M$ treatment variables or treatment components). In the dataset, there are $N$ study units. Let $Z_{n,m}$ denote the $m$-th observed treatment variable for unit $n$, where $n=1,\dots, N$ and $m=1,\dots,M$. Then, we define the observed treatment vector for unit $n$ as $\mathbf{Z}_{n}=(Z_{n,1},\cdots,Z_{n,M})$. Let $\mathcal{A}_{m}$ represent the collection of all possible values for the $m$-th treatment variable; then these values are either all distinct (e.g., in the continuous treatment case) or may have ties (e.g., in the ordinal or binary treatment case). Next, let the Cartesian product $\mathcal{A}=\mathcal{A}_{1}\times \dots \times \mathcal{A}_{M}$ represent the space of all possible treatment dose vectors in the study. If all the treatments are binary, the cardinality of $\mathcal{A}$ will be $2^M$. If all treatment variables take values in $\mathbb{R}$, then the treatment space $\mathcal{A}$ is $\mathbbm{R}^{M}$. Let $Y_n$ denote the observed outcome for unit $n$. Following the potential outcomes framework and assuming no interference among units (\citealp{neyman1923application, rubin1974estimating, rosenbaum2002observational,rosenbaum2020design}), we denote the potential outcome under the treatment vector $\mathbf{z}_n=(z_{n,1},\cdots,z_{n,M}) \in \mathcal{A}$ as $Y_n(\mathbf{z}_n)=Y_n(z_{n,1},\cdots,z_{n,M})$. Therefore, the observed outcome $Y_n=Y_n(\mathbf{z_n})$ when $\mathbf{Z}_n=\mathbf{z}_n$.

After clarifying the potential outcomes under factorial treatments, we describe the existing approaches to factorial matched observational studies and their limitations.

\begin{itemize}
    \item \textbf{Approach 1 (Separate Inferences):} A widely used approach is to match separately on each treatment variable \citep{rosenbaum2002observational, rosenbaum2020design}. This approach performs matching on one treatment variable while treating others as fixed, baseline covariates. However, this approach fails to define potential outcomes jointly with respect to all treatments. For example, in a study with two treatment variables, if units are matched only on one treatment variable (e.g., $Z_1$) while treating the other (e.g., $Z_2$) as fixed, then the potential outcomes for unit $n$ that we can infer is $Y_n(z_{n,1}, Z_{n,2})$, where $z_{n,1}\in \mathcal{A}_{1}$ and $Z_{n,2}$ is fixed. Conversely, if matched only on $Z_2$, the potential outcome we can learn about for unit $n$ is $Y_n(Z_{n,1}, z_{n,2})$, where $Z_{n,1}$ is fixed and $z_{n,2}\in \mathcal{A}_{2}$. As a result, there is no variation or source of randomization in the treatment variables that we treat as fixed, baseline covariates, and do not match on, which limits the ability to investigate their causal effect or interaction with the matched treatment.
    
    \item \textbf{Approach 2 (Dichotomization):} Another common strategy is to dichotomize each non-binary (e.g., continuous) treatment variable into a binary variable and apply matching methods for binary treatments \citep{fong2018continuous,Boyd2010sex,Nielsen2011aid,fogarty2019biased,yu2023matching}. This approach leads to a loss of information and conceptual ambiguity because dichotomizing non-binary treatment variables eliminates the clear correspondence between treatment doses and their associated potential outcomes. For instance, suppose the study includes a single non-binary treatment variable $Z_{n,1}$ for unit $n$, which is dichotomized into a binary variable $Z^*_{n,1}$ based on some ad hoc threshold $\gamma$, such that $Z^*_{n,1}=1$ if the observed value of $Z_{n,1}$ is at least $\gamma$, and $Z^*_{n,1}=0$ otherwise. Under this dichotomization, any two units with treatment levels $Z_{n,1}=z''$ and $Z_{n,1}=z'$, respectively, such that $z''>z'\geq \gamma$, are assigned to the same new binary treatment group. As a result, their potential outcomes are assumed to be identical, i.e., $Y_n(z'')=Y_n(z')$. This simplification discards the variation in potential outcomes within the same dichotomized or coarsened group.

    \item \textbf{Approach 3 (Aggregation:)} The third approach is to aggregate multiple, possible non-binary (e.g., continuous) treatments into a single composite score \citep{zhang2023social,frazier2024bias,pan2020sdi,zhang2021platform}. In this approach, researchers assign some prespecified weights to the original treatments to construct a combined treatment variable, effectively reducing the factorial treatment to a single dimension. However, this dimension reduction process relies on a strong assumption of a linear (or non-linear) combination of multiple treatment variables, which is impractical in the study. For example, under the linear combination assumption, the researchers assume that $Y_n(z_{n,1},\dots,z_{n,M})=h_{n}(\alpha_1z_{n,1}+\dots+\alpha_{M}z_{n,M})$ for each subject $n$, where $h_{n}$ is some unknown function and $\alpha_1,\dots,\alpha_M$ are some prespecified weights for the factorial treatments. However, it is possible that the treatment variables follow nonlinear combinations. For example, $Y_n(z_{n,1},\dots,z_{n,M})=\alpha_1z_{n,1}^2+\dots+\alpha_{M}z_{n,M}^2$ for subject $n$. In this case, the linear combination assumption can be violated, and the subsequent estimation and inference can be problematic. 

\end{itemize}

These limitations motivate us to develop a new framework for factorial matched observational studies that operates directly in the factorial treatment space and enables the definition of meaningful causal estimands aligned with these treatment variables. In the following section, we present our proposed design-based approach to address these methodological gaps.

\section{The Proposed Study Design: Factorial Matching with General Treatment Types and Its Associated Causal Estimands}\label{sec: design}

\subsection{A Universal Matching Design for Factorial Observational Studies with General Treatment Types}\label{sec: matching}

To focus on illustrating our main ideas without introducing complicated notations, in Sections  \ref{sec: design} and \ref{sec: inference}, we consider the bivariate treatment cases (i.e., two intervention factors with $M=2$). Our framework can be readily generalized to higher treatment dimensions such as $M\geq 3$.

The presence of two possibly non-binary treatment variables introduces several key considerations for the design stage. Firstly, as in the case of a single treatment variable, matching must achieve covariate balance, that is, units with similar observed covariates should be matched in the same set to control for confounders. Secondly, different combinations of treatment levels following the idea of non-bipartite matching for one continuous treatment \citep{lu2001matching,lu2011optimal,greevy2023nonbipartite,zhang2023matching}, the estimation of main effects requires constructing pairs of units that differ substantially in one treatment variable (for improving power of detecting treatment effects), while holding the other treatment approximately constant (to make the estimand more interpretable). This strategy should be applied symmetrically to both treatment variables.


To incorporate these design considerations, we propose a two-stage non-bipartite matching procedure that constructs matched sets containing four units with distinct treatment dose combinations while maintaining covariate similarity. After matching, each stratum consists of units with similar observed covariates but distinctly different patterns of treatment doses, thereby facilitating the estimation of both main and interaction effects (see Section \ref{sec: estimand} for details). The terms "low" and "high" refer to relatively lower or higher values of each treatment variable compared to those of other units within the same stratum, rather than binary indicators resulting from dichotomization. To construct such strata, our algorithm follows a two-stage matching procedure. First, we match two units with similar covariates as pairs based on one treatment variable. Next, treating each matched pair as a new unit, we optimally match two pairs together to form a stratum of four units (i.e., pairs of pairs) with treatment patterns as described in Table~\ref{table: patterns}. The procedure is formally described as follows:

\begin{table}[htbp]
\centering
\caption{Patterns of treatment doses for units in one stratum.}
\small
\vspace{-0.3cm}
\begin{tabular}{c|c c}
Treatment doses & $Z_{1}$ & $Z_{2}$ \\
\hline 
Unit 1 & Low &  Low \\
Unit 2 & Low &  High \\
Unit 3 & High & Low \\
Unit 4 & High & High \\
\end{tabular}
\label{table: patterns}
\end{table}

\textbf{Step 1}: We first conduct an optimal non-bipartite matching algorithm \citep{lu2001matching, lu2011optimal, baiocchi2012near, greevy2023nonbipartite, frazier2024bias}, also referred to as the "near-far" matching algorithm, to pair two units together based on the first treatment variable $Z_1$. The goal of this step is to pair units with the same or similar observed covariates but very different $Z_1$. To this end, let $n$ and $n'$ denote two distinct subjects, and define $\delta(\mathbf{x}_n,\mathbf{x}_{n'})\geq0$ as the distance between their observed covariates $\mathbf{x}_n$ and $\mathbf{x}_{n'}$, where $\delta(\mathbf{x}_n,\mathbf{x}_{n'})$ can be specified as the Mahalanobis distance or its rank-transformed version, and so on. Next, consider the following distance between two subjects:
\begin{equation*}
d(n,n')=\frac{\delta(\mathbf{x}_{n},\mathbf{x}_{n'})}{(Z_{n1}-Z_{n',1})^2}.
\end{equation*}
Specifically, in cases where $Z_{n,1}=Z_{n',1}$, $d(n,n')$ is set as $\infty$, regardless of the covariate similarity (i.e., even if $\delta(\mathbf{x}_n,\mathbf{x}_{n'})=0$). This ensures that units with identical first treatment doses are never matched together (otherwise, these pairs are concordant in treatment values and will not contribute any useful information for our statistical inference). In settings where treatment variables are binary, this non-bipartite matching algorithm will collapse to a bipartite matching algorithm using $\delta(\mathbf{x}_n,\mathbf{x}_{n'})$ (see Remark 2 for details).
       
The final set of matched pairs is obtained by minimizing the total distance across all selected pairs, subject to the constraint that each unit is matched exactly once. Consider a graph $G=(V,E)$, where the vertex set is defined as $V=\{1\dots,N\}$, representing the units in the study sample before matching. The edge set is given by $E=\{(n,n'):n,n'\in V,n<n'\}$, representing the set of matched pairs after matching. Notably, we impose a constraint that all units are matched without replacement, meaning no unit can be included in more than one matched pair. Therefore, if $(n,n') \in E$, then for all $n''\neq n'$, we have $(n,n'')\notin E$. Let $\mathcal{E}$ denote the collection of all $E$ that satisfy this constraint. The optimal set of matched pairs $E^* \in \mathcal{E}$ is obtained by solving the following optimization problem:
\vspace{-0.2in}
\begin{equation*}
    E^*=\arg\min_{E\in\mathcal{E}}\sum_{(n,n')\in E}d(n,n').
\end{equation*}  
\vspace{-0.4in}

\textbf{Step 2}: Then, we aim at constructing strata consisting of four units (i.e., matched quadruples) that follow the treatment patterns shown in Table~\ref{table: patterns}, while ensuring that these four units have similar observed covariates. To achieve this goal, we conduct a second-stage optimal non-bipartite matching algorithm to match two pairs together. Suppose there are $Q$ pairs obtained from Step 1. For each pair $q \in \{1,\dots,Q\}$, we order the two units so that the first subject receives a higher dose of the first treatment variable. That is, we assigns labels $q1$ and $q2$ to the two subjects in pair $q$ so that $Z_{q1,1}>Z_{q2,1}$.
    
Let $q$ and $q'$ denote two distinct pairs. We then order the four subjects from these two pairs by their values of the first treatment variable $Z_1$. Among the two subjects with lower $Z_1$, the one with the lower $Z_2$ is labeled as "low" for the second treatment, while the other is labeled as "high." Similarly, we assign "low" and "high" labels for the second treatment among the two subjects with higher $Z_1$. Next, we consider the following distance between two pairs: 
\begin{align*}
&d_2(q,q')=\min\Bigg\{\Bigg(\frac{\delta(\mathbf{x}_{q1},\mathbf{x}_{q'1})+{(Z_{q1,1}-Z_{q'1,1})^2}}{(Z_{q1,2}-Z_{q'1,2})^2}+\frac{\delta(\mathbf{x}_{q2},\mathbf{x}_{q'2})+{(Z_{q2,1}-Z_{q'2,1})^2}}{(Z_{q2,2}-Z_{q'2,2})^2}\Bigg), \\ &\Bigg(\frac{\delta(\mathbf{x}_{q1},\mathbf{x}_{q'2})+{(Z_{q1,1}-Z_{q'2,1})^2}}{(Z_{q1,2}-Z_{q'2,2})^2}+\frac{\delta(\mathbf{x}_{q2},\mathbf{x}_{q'1})+{(Z_{q2,1}-Z_{q'1,1})^2}}{(Z_{q2,2}-Z_{q'1,2})^2}\Bigg)\Bigg\}+W\times\mathbbm{1}\big\{B_1\vee B_2 \vee B_3 \big\}.
\end{align*}
For any potential pairing of two first-stage pairs, we minimize the sum of two pairwise distances. In each distance, the numerator contains differences in the first treatment and covariates (favoring smaller differences), and the denominator contains differences in the second treatment (favoring larger differences). Because the first-stage matching has already constructed pairs in which the two units differ substantially in the first treatment, pairing two such pairs together naturally yields a quadruple with two “high” and two “low” first-treatment doses. In the second stage, to achieve the treatment-dose patterns required by Table~\ref{table: patterns}, we further encourage the two pairs being matched to have similar covariates and similar first-treatment doses but substantially different second-treatment doses. This combination helps ensure that each resulting stratum consists of four covariate-balanced units arranged in the desired treatment-dose configuration.

In addition, a penalty term, $W\times\mathbbm{1}\{B_1\vee B_2 \vee B_3\}$, is added to $d_2(q,q')$. This term consists of two components: the first is the penalty parameter $W$, which is a strictly positive and relatively large number (e.g., 5000); the second is an indicator function that activates the penalty. Specifically, if any of the following conditions are satisfied, the penalty $W$ is added to $d_2(q,q')$:
\begin{itemize}
    \item $B_1$: The absolute difference between the "high" (or "low") values of $Z_1$ across two pairs is larger than a threshold (e.g., 0.05). This condition is penalized because large discrepancies in the “high” or “low” values reduce comparability within strata and complicate the interpretation of treatment effects.
    \item $B_2$: The absolute difference between the "high" (or "low") values of $Z_2$ across two pairs is larger than a threshold (e.g., 0.05). This condition is penalized as the same reason as $B_1$.
    \item $B_3$: The second treatment values for two subjects with "high" $Z_1$ are either both smaller or both larger than those of the two subjects with "low" $Z_1$. When this occurs, the resulting stratum contains only two treatment patterns (e.g., low-low and high-high or low-high and high-low), which limits the ability to estimate and conduct valid inference on both main and interaction effects (see Section~\ref{sec: inference} for details).
        
\end{itemize}

If the treatment variables are all binary in the study, two treated-control pairs are matched to construct a stratum of four units that collectively exhibit all four treatment combinations: (1,1), (1,0), (0,1), and (0,0).
   
Finally, we construct a new graph $G^{(2)}=(V^{(2)},E^{(2)})$, where $V^{(2)}=\{1,\dots,Q\}$ represents the set of matched pairs from the first stage. Edges $E^{(2)}=\{(q,q'):q,q'\in V^{(2)}, q<q'\}$ correspond to potential strata of four units. As before, two constraints must be considered for edges in $E^{(2)}$: 1) no individual should be matched with itself; and 2) matching must be done without replacement. Let $\mathcal{E}^{(2)}$ denote the collection of all valid edge subsets. The optimal set of strata is obtained by solving:
\vspace{-0.1in}
\begin{equation*}
E^{(2)*}=\arg\min_{E^{(2)}\in\mathcal{E}^{(2)}}\sum_{(q,q')\in E^{(2)}}d_2(q,q').
\end{equation*}  

Notably, the first-stage matching may differ if it is constructed using the second treatment variable $Z_2$, which in turn changes the overall construction of strata. In this case, label assignments within each pair are based on $Z_2$, such that $Z_{q1,2}>Z_{q2,2}$. Consequently, the distance formula $d_2(q,q')$ in Step 2 would be modified: the roles of $Z_1$ and $Z_2$ are reversed, with squared differences in $Z_1$ appearing in the denominator and squared differences in $Z_2$ in the numerator. The conditions for applying the penalty term are also updated accordingly. To obtain the best matching result, we may apply the procedure twice: once for matching on $Z_1$ in Step 1, and once for matching on $Z_2$ in Step 1. Then, the version that yields the smaller total distance in Step 2 is selected.

After matching, for each stratum of four units, we assign labels based on treatment values. Units are first ordered by $Z_1$. Among the two units with comparatively lower $Z_1$ values, the one with the lower $Z_2$ value is labeled as $(l,l)$, and the other is labeled as $(l,h)$. Next, among the two subjects with higher $Z_1$ values, the one with the lower $Z_2$ value is labeled as $(h,l)$, and the other as $(h,h)$. Table~\ref{table: example} presents an example stratum. In this example, Unit 1 and Unit 2 have relatively low $Z_1$ values. Between them, Unit 1 has the lower $Z_2$ value and is thus labeled as $(l,l)$, while Unit 2 is labeled as $(l,h)$. Similarly, we can label Unit 3 as $(h,l)$ and Unit 4 as $(h,h)$. Next, we can conduct randomization-based inference in the following two ways: (i) investigate a parametric dose-response relationship between treatments and the outcome (Section \ref{sec: fisher}), or (ii) apply a non-parametric approach to estimate the sample average main and interaction effects (Section \ref{sec: neyman}).

\begin{table}[htbp]
\centering
\caption{Example of treatment dose patterns in a matched stratum.}
\small
\vspace{-0.3cm}
\begin{tabular}{l|c c}
Treatment doses & $Z_{1}$ & $Z_{2}$ \\
\hline 
Unit 1 $(l,l)$ & 0.31 &  0.30 \\
Unit 2 $(l,h)$ & 0.35 &  0.82 \\
Unit 3 $(h,l)$ & 0.74 & 0.23 \\
Unit 4 $(h,h)$ & 0.86 & 0.71 \\
\end{tabular}
\label{table: example}
\end{table}

\begin{remark}
While the ideal scenario is to construct matched sets containing four units with distinct combinations of treatment doses, this is not always feasible in practice. In some studies, the total number of units may not be a multiple of four, or it may be challenging to match four units with sufficiently similar covariates and distinct treatment patterns together. In such cases, matched strata containing only two or three units are still retained, as they still can contribute meaningful information to the subsequent estimation and inference. 

\end{remark}

\begin{remark}
When all treatment variables in the study are binary, for instance, if $Z_{k,1} \in \{0,1\}$ indicates treatment status (1 for treated and 0 for control), then $d(k,k')=\infty$ if $Z_{k,1}=Z_{k',1}$, and $d(k,k')=\delta(\mathbf{x}_k,\mathbf{x}_{k'})$ otherwise. In this case, Step 1 naturally reduces to a bipartite matching algorithm that pairs treated and control units based on covariate distance $\delta(\mathbf{x}_k,\mathbf{x}_{k'})$.
\end{remark}

\subsection{The Proposed Causal Estimands: Generalized Factorial Neyman-Type Estimands for Main and Interaction Effects} \label{sec: estimand}

Suppose there are $I$ matched sets formed using the quadruple matching design introduced in Section~\ref{sec: matching}, with each matched set $i \ (i=1,\dots,I)$ containing four units. This yields a total of $4I$ units in the dataset. For unit $j \in \{1,2,3,4\}$ within matched set $i$, let $Z_{ij,1}$ and $Z_{ij,2}$ denote the first and second observed treatment variables, let $Y_{ij}$ denote the observed outcome, and let $\mathbf{x}_{ij}=(x_{ij,1},\dots,x_{ij,K})$ denote the K-dimensional vector of observed covariates. Define the vector of all observed outcomes in the dataset as $\mathbf{Y}=(Y_{11},\dots,Y_{I4})$. For unit $ij$, define the observed treatment vector as $\mathbf{Z}_{ij}=(Z_{ij,1},Z_{ij,2})$, and let $\mathbf{Z}=(\mathbf{Z}_{11}, \dots, \mathbf{Z}_{I4})$ denote the collection of all observed treatment vectors across all units. Let $\mathcal{A}_1$ and $\mathcal{A}_2$ denote the sets of all possible values for the first and second treatment variable, and let $\mathcal{A}=\mathcal{A}_1 \times \mathcal{A}_2$ denote the space of all possible treatment vectors. Following the potential outcomes framework \citep{neyman1923application,rubin1974estimating,rosenbaum1989optimal}, we let $Y_{ij}(\mathbf{z}_{ij})=Y_{ij}(z_{ij,1},z_{ij,2})$ denote the potential outcome under the treatment vector $\mathbf{z}_{ij}=(z_{ij,1},z_{ij,2}) \in \mathcal{A}$. Under the design-based (randomization-based) inference framework, all the potential outcomes $Y_{ij}(\mathbf{z}_{ij})$ are treated as fixed, and the only source of randomness that enters into inference is the random assignment of treatment values after matching \citep{rosenbaum1989sensitivity,rosenbaum2002observational, gastwirth1998dual,baiocchi2010building,zhang2023social, zhang2023matching}.

After matching, four units with similar covariates but different treatment patterns are matched in a stratum; each unit is labeled according to the relative levels of its treatment variables. Specifically, within stratum $i$, let $(z_{i,1}(l,h), z_{i,2}(l,h))$ denote the treatment vector for the unit labeled as $(l,h)$, where $l$ and $h$ indicate relatively low and high values of the first and second treatment respectively. Similarly, the treatment vectors for the other labeled units are denotes as $(z_{i,1}(l,l),z_{i,2}(l,l))$, $(z_{i,1}(h,h),z_{i,2}(h,h))$ and $(z_{i,1}(h,l),z_{i,2}(h,l))$. 

\begin{figure}[b!]
\centering
\begin{tikzpicture}[item/.style={anchor=west}]

\node (t1) at (0,3) {$(z_{i,1}(h,h),\ z_{i,2}(h,h))$};
\node (t2) [below=0.5cm of t1] {$(z_{i,1}(h,l),\ z_{i,2}(h,l))$};
\node (t3) [below=0.5cm of t2] {$(z_{i,1}(l,h),\ z_{i,2}(l,h))$};
\node (t4) [below=0.5cm of t3] {$(z_{i,1}(l,l),\ z_{i,2}(l,l))$};

\node (u1) at (6,3) {Unit $i_1$};
\node (u2) [below=0.5cm of u1] {Unit $i_2$};
\node (u3) [below=0.5cm of u2] {Unit $i_3$};
\node (u4) [below=0.5cm of u3] {Unit $i_4$};

\draw (t1.east) -- (u3.west);
\draw (t2.east) -- (u1.west);
\draw (t3.east) -- (u4.west);
\draw (t4.east) -- (u2.west);

\end{tikzpicture}
\caption{Treatment labels of four units in each matched set.}
\end{figure}
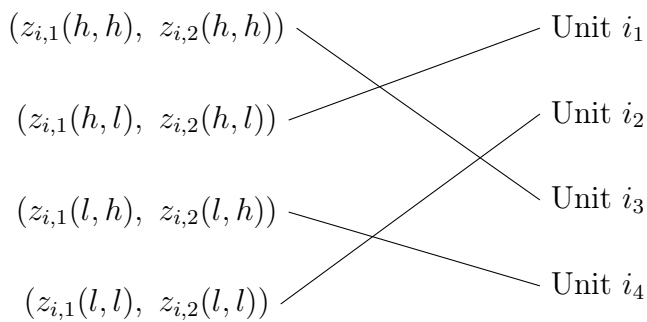

For each unit $ij$ in the study, we can define the following four potential outcomes:
\vspace{-0.1in}
\begin{equation}
\begin{split}
    &Y_{ij}(l,l)\overset{\Delta}{=}Y_{ij}(z_{i,1}(l,l),z_{i,2}(l,l)), \,\, Y_{ij}(l,h)\overset{\Delta}{=}Y_{ij}(z_{i,1}(l,h),z_{i,2}(l,h)),  \\       
    &Y_{ij}(h,l)\overset{\Delta}{=}Y_{ij}(z_{i,1}(h,l),z_{i,2}(h,l)),\,\, Y_{ij}(h,h)\overset{\Delta}{=}Y_{ij}(z_{i,1}(h,h),z_{i,2}(h,h)).
\end{split}
\label{formula: potential}
\end{equation}
For example, refer back to the example stratum shown in Table~\ref{table: example} in Section \ref{sec: matching}. For unit 1 in the stratum, the four potential outcomes are: $Y_{i1}(l,l)\overset{\Delta}{=}Y_{i1}(0.31,0.30)$, $Y_{i1}(l,h)\overset{\Delta}{=}Y_{i1}(0.35,0.82)$, $Y_{i1}(h,l)\overset{\Delta}{=}Y_{i1}(0.74,0.23)$, and $Y_{i1}(h,h)\overset{\Delta}{=}Y_{i1}(0.86,0.71)$.

In each matched set $i$, given the treatment dose set $\mathbf{z}_i=\big\{(z_{i,1}(l,h), z_{i,2}(l,h)),(z_{i,1}(l,l),z_{i,2}(l,l)), \\ (z_{i,1}(h,h),z_{i,2}(h,h)),(z_{i,1}(h,l),z_{i,2}(h,l))\big\}$, corresponding to $4!$ permutations of $\mathbf{z}_i$, there are $4!$ possible realizations of the observed treatment dose set $\mathbf{Z}_i=(\mathbf{Z}_{i1},\mathbf{Z}_{i2},\mathbf{Z}_{i3},\mathbf{Z}_{i4})$ (i.e., there are $4!$ different ways to assign four different treatment dose vectors to four units). Then, the space of all $4!$ permutations of $\mathbf{z}_i$ can be denoted by $\mathcal{Z}_{i}$, where $|\mathcal{Z}_i|=4!$. Let $\mathcal{Z}=\mathcal{Z}_1\times\mathcal{Z}_2\times\dots\times\mathcal{Z}_I$ denote all possible dose assignments in randomization inference for the whole matched dataset. The cardinality of $\mathcal{Z}=(4!)^I$. Under the no unmeasured confounding and the ignorability assumptions, the treatment dose assignments are uniformly distributed within each matched set. See Section~\ref{sec: fisher} for details.

In factorial designs with two binary factors, widely considered causal estimands are the main effects for each factor and the interaction effect between them. The main effect can be defined by contrasting, for each unit, one half of the potential outcomes with the other half of the potential outcomes. For example, the main effect of factor 1 corresponds to the marginal effect of $Z_1$ averaged over the levels of $Z_2$, and can be calculated as the difference between the average potential outcomes when $Z_1$ is at its high level versus its low level. Next, the interaction effect between factor 1 and factor 2 captures how the effect of one treatment variable depends on the level of the other. It can be expressed as the difference between two conditional effects. For instance, to assess how the effect of $Z_1$ depends on the level of $Z_2$, we compute the difference in the effect of $Z_1$ when $Z_2$ is at its high level versus when it is at its low level. In our work, the treatment variables are not restricted to being binary; they may also be ordinal or continuous. Therefore, we propose new estimands, referred to as generalized factorial Neyman-type estimands, that extend the classical definitions of main and interaction effects to accommodate general treatment types. 

To rigorously define the generalized factorial Neyman-type estimands, we first introduce some notations. For each unit $ij$, let $\overline{Y}_{ij}(h,\cdot)=\frac{1}{2}\big\{Y_{ij}(h,h)+Y_{ij}(h,l)\big\}$ and $\overline{Y}_{ij}(l,\cdot)=\frac{1}{2}\big\{Y_{ij}(l,h)+Y_{ij}(l,l)\big\}$ denote the average potential outcomes when $Z_1$ is at high and low levels, respectively. Next, for each stratum $i$, define $\overline{z}_{i}(h,\cdot)$ as the average value of $z_{i,1}(h,h)$ and $z_{i,1}(h,l)$, and $\overline{z}_{i}(l,\cdot)$ as the average value of $z_{i,1}(l,h)$ and $z_{i,1}(l,l)$. Analogous notations $\overline{Y}_{ij}(\cdot,h)$, $\overline{Y}_{ij}(\cdot,l)$, $\overline{z}_{i}(\cdot,h)$, and $\overline{z}_{i}(\cdot,l)$ are defined similarly for the second treatment variable $Z_2$. Then, for each unit $ij$, we define the main effect of the first treatment variable $Z_1$ as the difference between 
$\overline{Y}_{ij}(h,\cdot)$ and $\overline{Y}_{ij}(l,\cdot)$. When $Z_1$ is continuous, this difference must be scaled by the magnitude of the change in $Z_1$. Then, the unit-level main effect of $Z_1$ for each unit $ij$ is defined as $\tau_{ij,1}=\frac{\overline{Y}_{ij}(h,\cdot)-\overline{Y}_{ij}(l,\cdot)}{\overline{z}_{i}(h,\cdot)-\overline{z}_{i}(l,\cdot)}$. The average of these $4I$ unit-level main effects of $Z_1$ across all $4I$ units defines the finite-population factorial main effect estimand for $Z_1$:
\vspace{-0.7cm}
\begin{equation}\label{eqn:main1}
\begin{split}
    \tau_{1}=\frac{1}{4I}\sum_{i=1}^{I}\sum_{j=1}^{4}\frac{\overline{Y}_{ij}(h,\cdot)-\overline{Y}_{ij}(l,\cdot)}{\overline{z}_{i}(h,\cdot)-\overline{z}_{i}(l,\cdot)}. 
\end{split}
\end{equation}
Similarly, the finite-population factorial main effect estimand for $Z_2$ is denoted as:
\vspace{-0.2cm}
\begin{equation}\label{eqn:main2}
\begin{split}
    \tau_{2}=\frac{1}{4I}\sum_{i=1}^{I}\sum_{j=1}^{4}\frac{\overline{Y}_{ij}(\cdot,h)-\overline{Y}_{ij}(\cdot,l)}{\overline{z}_{i}(\cdot,h)-\overline{z}_{i}(\cdot,l)}. 
\end{split}
\end{equation}

Next, we define the conditional effect of $Z_1$ when $Z_2$ is at its high and low level, respectively, as 
\vspace{-0.2in}
\begin{equation*}
    \lambda_{ij,1}(h)=\frac{Y_{ij}(h,h)-Y_{ij}(l,h)}{z_{i,1}(h,h)-z_{i,1}(l,h)}  \ \  \text{and} \ \
    \lambda_{ij,1}(l)=\frac{Y_{ij}(h,l)-Y_{ij}(l,l)}{z_{i,1}(h,l)-z_{i,1}(l,l)}.
\end{equation*}
Similarly, the conditional effect of $Z_2$ can be defined when $Z_1$ is at its high and low level as
\begin{equation*}
    \lambda_{ij,2}(h)=\frac{Y_{ij}(h,h)-Y_{ij}(h,l)}{z_{i,2}(h,h)-z_{i,2}(h,l)} \ \  \text{and} \ \ \lambda_{ij,2}(l)=\frac{Y_{ij}(l,h)-Y_{ij}(l,l)}{z_{i,2}(l,h)-z_{i,2}(l,l)}.
\end{equation*}

\begin{remark}
When calculating these conditional effects within each matched set, the high and low values of the conditional treatment variables may not be exactly the same in the set. If these values differ substantially, the interpretability of the conditional effect may be compromised. For example, if the "high" values of $Z_2$ in matched set $i$ differ substantially (e.g., 0.5 vs. 0.9), then interpreting $\lambda_{ij,1}(h)$ becomes problematic. In this case, the observed difference $Y_{ij}(h,h)-Y_{ij}(l,h)$ may be due to not only the effect of changing $Z_1$ from low to high, but also the variation in $Z_2$, thus confounding the interpretation of the condition effect. To mitigate this issue, we impose penalty conditions when calculating $d_2(q,q')$ within the matching procedure to ensure that the high and low values of both $Z_1$ and $Z_2$ are as similar as possible in each matched set (see Section \ref{sec: matching} for details). 
\end{remark}

Then, we can define the unit-level two-way interaction effect between the two treatment variables. For each unit $ij$, to measure how $Z_2$ modifies the effect of $Z_1$, we compute the difference between the conditional effects of $Z_1$ when $Z_2$ is at its high level versus its low level. When $Z_2$ is continuous, the effect of $Z_1$ must be scaled by the magnitude of the change in $Z_2$. Similarly, when $Z_1$ is also continuous, the interaction effect should be scaled by the change in $Z_1$, which we compute using the average high and low values of $Z_1$ within the matched set. The formula for the unit-level interaction effect, which quantifies how the effect of $Z_1$ changes depending on the level of $Z_2$ for each unit $ij$, is given as $\tau_{ij,3}=\frac{\lambda_{ij,1}(h)-\lambda_{ij,1}(l)}{\overline{z}_{i}(\cdot,h)-\overline{z}_{i}(\cdot,l)}$.
Therefore we denote the finite-population factorial interaction effect $\tau_3$ as the average of the $4I$ unit-level interaction effects $\tau_{ij,3}$, each of which measures how the effect of $Z_1$ varies depending on the level of $Z_2$:
\vspace{-0.4cm}
\begin{equation}\label{eqn:inter1}
\begin{split}
    \tau_{3}&=\frac{1}{4I}\sum_{i=1}^{I}\sum_{j=1}^{4}\frac{\lambda_{ij,1}(h)-\lambda_{ij,1}(l)}{\overline{z}_{i}(\cdot,h)-\overline{z}_{i}(\cdot,l)}.
\end{split}
\end{equation}

Likewise, the average of the $4I$ unit-level interaction effects, each measuring how the effect of $Z_2$ changes with the level of $Z_1$, defines the finite-population factorial interaction effect $\tau_4$:
\vspace{-0.4cm}
\begin{equation}\label{eqn:inter2}
\begin{split}
    \tau_{4}&=\frac{1}{4I}\sum_{i=1}^{I}\sum_{j=1}^{4}\frac{\lambda_{ij,2}(h)-\lambda_{ij,2}(l)}{\overline{z}_{i}(h,\cdot)-\overline{z}_{i}(l,\cdot)}.
\end{split}
\end{equation}
When the treatment variables are binary, our proposed factorial interaction effect estimands $\tau_{3}$ and $\tau_{4}$ reduce to the classic finite-population factorial interaction effect estimands introduced in \citet{dasgupta2015factorial}.

\section{Randomization-Based Causal Inference }\label{sec: inference}

\subsection{Fisher Randomization Tests for Dose-Response Relationship}\label{sec: fisher}

Let $\mathcal{H}=\{(\mathbf{x}_{ij},Y_{ij}(l,l),Y_{ij}(l,h),Y_{ij}(h,l),Y_{ij}(h,h)):i=1,\dots,I,j=1,2,3,4\}$ denote the collection of all observed covariates and potential outcomes for all units in the dataset. In the design-based inference framework, we condition on the observed covariates, the potential outcomes, and the full set of possible treatment dose assignments for the matched dataset. Causal conclusions are then drawn solely based on the randomness induced by treatment assignments. Conditional on matching and the ignorability assumption, the treatment dose assignments are completely random within each matched set $i$. Then, under independence across matched sets, the joint assignment probability across all sets is: 
\vspace{-0.1in}
\begin{equation*}
    P(\mathbf{Z}=\mathbf{z}|\mathcal{H},\mathcal{Z})=\Big(\frac{1}{4!}\Big)^I \ \text{for \ all} \ \mathbf{z} \in \mathcal{Z}.
\end{equation*}

Next, we can test the dose-response relationship for unit $ij$ between the factorial treatment variables and the potential outcomes. Let $(z_1^*,z_2^*)$ be the arbitrary reference doses, then the following null hypothesis can represent the dose-response relationship for all $i, j$: 
\vspace{-0.3cm}
\begin{equation*}
    H_{0}^{\text{sharp}}:Y_{ij}(z_{1},z_{2})-Y_{ij}(z_1^*,z_2^*)=f((z_{1},z_{2});(z_1^*,z_2^*), \mathbf{x}_{ij},\boldsymbol{\theta}) \text{ for all ($z_{1}, z_{2}$)}.
\end{equation*}
Here, $f$ is some prespecified function for describing a structured dose-response relationship, and $\boldsymbol{\theta}$ are some prespecified parameters to be tested in that structured dose-response relationship. For example, we can test the null hypothesis of no effect by imposing $Y_{ij}(z_1,z_2)=Y_{ij}(z_1^*,z_2^*)$, which implies no difference in potential outcomes across any two treatment-dose combinations. A linear specification can also be tested. For example, we can test for $Y_{ij}(z_1,z_2)-Y_{ij}(z_1^*,z_2^*)=\theta_1(z_1-z_1^*)+\theta_2(z_2-z_2^*)$. Beyond linearity, we can test for nonlinear relationships or include interaction terms in the hypothesis, such as the interaction between two treatments or between a treatment and a covariate (e.g., $Y_{ij}(z_1,z_2)-Y_{ij}(z_1^*,z_2^*)=\theta_1(z_1-z_1^*)+\theta_2(z_2-z_2^*)+\theta_3x_{ij,1}(z_1-z_1^*)$). Moreover, a Single Index Model (SIM) \citep{chen2019sim, jennifer2015sim} can be used to model unit-level causal effects under the sharp null hypothesis: $Y_{ij}(z_1,z_2)-Y_{ij}(z_1^*,z_2^*)=f(\theta_1(z_1-z_1^*)+\theta_2(z_2-z_2^*)+\theta_3(z_1z_2-z_1^*z_2^*))$ for some function $f$, where $\theta_1,\theta_2,\theta_3\in[0,1]$ and $\theta_1+\theta_2+\theta_3=1$.

Given by a test statistic $T(\mathbf{Z},\mathbf{Y})$ and its observed value $t$, the corresponding one-sided exact $p$-value can be calculated by:
\begin{equation*}
    P(T(\mathbf{Z},\mathbf{Y})\geq t|\mathcal{H},\mathcal{Z},H_{0}^{\text{sharp}})=\frac{|\{\mathbf{Z}\in\mathcal{Z}:T(\mathbf{Z},\mathbf{Y})\geq t\}|}{|\mathcal{Z}|},
\end{equation*}
which can be efficiently approximated by the Monte-Carlo method when the sample size is large in the study \citep{rosenbaum2002observational, imbens2015causal}. Then, a confidence region for $\boldsymbol{\theta}$ can be calculated by inverting the permutation test. See Appendix for details.

\subsection{Neyman-type Inference for the Proposed Causal Estimands}\label{sec: neyman}

In this section, we propose Neyman-type estimators for the generalized factorial Neyman-type estimands introduced in Section~\ref{sec: estimand}, which include two main effects and two interaction effects. Given the structural similarity among these estimands (see formulas~\ref{eqn:main1},~\ref{eqn:main2},~\ref{eqn:inter1},and~\ref{eqn:inter2}), we adopt a unified form for their estimators. Specifically, each estimator is defined as the average of stratum-level contributions across $I$ matched strata:
\vspace{-0.4cm}
\begin{equation*}
    \begin{split}
        \widehat{\tau}_{a}=\frac{1}{I}\sum_{i=1}^{I}V_i^{(a)}, \ \ \ \text{for } a \in \{1,2,3,4\},
    \end{split}
\end{equation*}
where $\widehat{\tau}_1$ and $\widehat{\tau}_2$ denote the estimators for the main effects of the first and second treatment variables $Z_{1}$ and $Z_{2}$, respectively, and $\widehat{\tau}_3$ and $\widehat{\tau}_4$ denote the two estimators for the two factorial interaction effects (one indicating the varying effects of $Z_{1}$ under different levels of $Z_{2}$, and one indicating the varying effects of $Z_{2}$ under different levels of $Z_{1}$). The stratum-level contributions $V_i^{(a)}$ vary by estimand and are defined as below:
\begin{itemize}
    \item Main effect of the first treatment variable (a=1):
    \begin{equation*}
    \begin{split}
       \hspace*{-2em}  V_i^{(1)}=\sum_{j=1}^{4}\bigg\{&\frac{1}{2}\frac{\mathbbm{1}\{Z_{ij,1}=z_{i,1}(h,h),Z_{ij,2}=z_{i,2}(h,h)\}Y_{ij}}{\Bar{z}_{i}(h,\cdot)-\Bar{z}_{i}(l,\cdot)}+\frac{1}{2}\frac{\mathbbm{1}\{Z_{ij,1}=z_{i,1}(h,l),Z_{ij,2}=z_{i,2}(h,l)\}Y_{ij}}{\Bar{z}_{i}(h,\cdot)-\Bar{z}_{i}(l,\cdot)} \\ 
        &-\frac{1}{2}\frac{\mathbbm{1}\{Z_{ij,1}=z_{i,1}(l,h),Z_{ij,2}=z_{i,2}(l,h)\}Y_{ij}}{\Bar{z}_{i}(h,\cdot)-\Bar{z}_{i}(l,\cdot)}-\frac{1}{2}\frac{\mathbbm{1}\{Z_{ij,1}=z_{i,1}(l,l),Z_{ij,2}=z_{i,2}(l,l)\}Y_{ij}}{\Bar{z}_{i}(h,\cdot)-\Bar{z}_{i}(l,\cdot)}\bigg\}.
    \end{split}
    \end{equation*}
    
    \item Main effect of the second treatment variable (a=2):
    \begin{equation*}
    \begin{split}
    \hspace*{-2em} 
        V_i^{(2)}=\sum_{j=1}^{4}\bigg\{&\frac{1}{2}\frac{\mathbbm{1}\{Z_{ij,1}=z_{i,1}(h,h),Z_{ij,2}=z_{i,2}(h,h)\}Y_{ij}}{\Bar{z}_{i}(\cdot,h)-\Bar{z}_{i}(\cdot,l)}+\frac{1}{2}\frac{\mathbbm{1}\{Z_{ij,1}=z_{i,1}(l,h),Z_{ij,2}=z_{i,2}(l,h)\}Y_{ij}}{\Bar{z}_{i}(\cdot,h)-\Bar{z}_{i}(\cdot,l)} \\ 
        &-\frac{1}{2}\frac{\mathbbm{1}\{Z_{ij,1}=z_{i,1}(h,l),Z_{ij,2}=z_{i,2}(h,l)\}Y_{ij}}{\Bar{z}_{i}(\cdot,h)-\Bar{z}_{i}(\cdot,l)}-\frac{1}{2}\frac{\mathbbm{1}\{Z_{ij,1}=z_{i,1}(l,l),Z_{ij,2}=z_{i,2}(l,l)\}Y_{ij}}{\Bar{z}_{i}(\cdot,h)-\Bar{z}_{i}(\cdot,l)}\bigg\}.
    \end{split}
    \end{equation*}
    \item Interaction effect that quantifies how the effect of $Z_1$ changes depending on the level of $Z_2$ (a=3):
    \begin{equation*}
    \begin{split}
    \hspace*{-2em}
        V_i^{(3)}=\sum_{j=1}^{4}\bigg\{&\frac{\mathbbm{1}\{Z_{ij,1}=z_{i,1}(h,h),Z_{ij,2}=z_{i,2}(h,h)\}Y_{ij}}{(z_{i,1}(h,h)-z_{i,1}(l,h))(\Bar{z}_{i}(\cdot,h)-\Bar{z}_{i}(\cdot,l))}-\frac{\mathbbm{1}\{Z_{ij,1}=z_{i,1}(l,h),Z_{ij,2}=z_{i,2}(l,h)\}Y_{ij}}{(z_{i,1}(h,h)-z_{i,1}(l,h))(\Bar{z}_{i}(\cdot,h)-\Bar{z}_{i}(\cdot,l))} \\ 
        &-\frac{\mathbbm{1}\{Z_{ij,1}=z_{i,1}(h,l),Z_{ij,2}=z_{i,2}(h,l)\}Y_{ij}}{(z_{i,1}(h,l)-z_{i,1}(l,l))(\Bar{z}_{i}(\cdot,h)-\Bar{z}_{i}(\cdot,l))}+\frac{\mathbbm{1}\{Z_{ij,1}=z_{i}(l,l),Z_{ij,2}=z_{i,2}(l,l)\}Y_{ij}}{(z_{i,1}(h,l)-z_{i,1}(l,l))(\Bar{z}_{i}(\cdot,h)-\Bar{z}_{i}(\cdot,l))}\bigg\}.
    \end{split}
    \end{equation*}
    \item Interaction effect that quantifies how the effect of $Z_2$ changes depending on the level of $Z_1$ (a=4):
    \begin{equation*}
    \begin{split}
    \hspace*{-2em}
        V_i^{(4)}=\sum_{j=1}^{4}\bigg\{&\frac{\mathbbm{1}\{Z_{ij,1}=z_{i,1}(h,h),Z_{ij,2}=z_{i,2}(h,h)\}Y_{ij}}{(z_{i,2}(h,h)-z_{i,2}(h,l))(\Bar{z}_{i}(h,\cdot)-\Bar{z}_{i}(l,\cdot))}-\frac{\mathbbm{1}\{Z_{ij,1}=z_{i,1}(h,l),Z_{ij,2}=z_{i,2}(h,l)\}Y_{ij}}{(z_{i,2}(h,h)-z_{i,2}(h,l))(\Bar{z}_{i}(h,\cdot)-\Bar{z}_{i}(l,\cdot))} \\ 
        &-\frac{\mathbbm{1}\{Z_{ij,1}=z_{i,1}(l,h),Z_{ij,2}=z_{i,2}(l,h)\}Y_{ij}}{(z_{i,2}(l,h)-z_{i,2}(l,l))(\Bar{z}_{i}(h,\cdot)-\Bar{z}_{i}(l,\cdot))}+\frac{\mathbbm{1}\{Z_{ij,1}=z_{i,1}(l,l),Z_{ij,2}=z_{i,2}(l,l)\}Y_{ij}}{(z_{i,2}(l,h)-z_{i,2}(l,l))(\Bar{z}_{i}(h,\cdot)-\Bar{z}_{i}(l,\cdot))}\bigg\}.
    \end{split}
    \end{equation*}
\end{itemize}

\begin{proposition}\label{prop:unbiased}
    Conditional on matching and the no unmeasured confounding assumption, we have $E(\widehat{\tau}_a|\mathcal{Z})=\tau_a$ for $a \in \{1,2,3,4\}$.
\end{proposition}

Proposition~\ref{prop:unbiased} shows that all four proposed estimators are unbiased for estimating the corresponding generalized factorial Neyman-type estimands. All the proofs are provided in the supplementary material.

Next, we can derive the asymptotically valid variance estimators for $\widehat{\tau}_a$, in which we also have the flexibility to incorporate the covariate information to improve efficiency. The key idea is to extend the existing design-based variance estimator for Neyman-type estimands in matched studies \citep{fogarty2018mitigating, zhang2024bridging, zhu2024bias, frazier2024bias} from the single treatment case to factorial treatments. Let $Q$ be any prespecified $I \times L$ matrix with $I>L$, where $I$ denotes the number of matched sets. For example, a vanilla choice for $Q$ is the $I\times 1$ vector of ones (i.e., a unit vector). More generally, to improve efficiency, $Q$ can incorporate covariate information aggregated at the matched set level. For example, when $K<I-1$, we can define $Q=(\mathbf{1}_{I\times1},\overline{\mathbf{x}}_1,\cdots,\overline{\mathbf{x}}_K)$, where $\mathbf{1}_{I\times1}=(1,\cdots,1)^T$ is an $I$-dimensional unit vector and each $\overline{\mathbf{x}}_k=(4^{-1}\sum_{j=1}^{4}x_{1jk},\cdots,4^{-1}\sum_{j=1}^{4}x_{Ijk})^{T}$ represents the mean value of the $k$-th covariate within each matched set. Let $H_Q=Q(Q^TQ)^{-1}Q^T$ be the hat matrix corresponding to $Q$, and let $h_{Qii}$ denote the $i$-th diagonal element of $H_Q$. Define $y_{i}^{(a)}=V_{i}^{(a)}/\sqrt{1-h_{Qii}}$ and $\mathbf{y}^{(a)}=(y_1^{(a)},\cdots,y_I^{(a)})$. Let $\mathcal{I}$ be the $I\times I$ identity matrix. Then, the variance of $\widehat{\tau}_a$ can be estimated using the following formula:
\begin{equation*}
    S^2(Q)_a=\frac{1}{I^2}\mathbf{y}^{(a)}(\mathcal{I}-H_Q)\mathbf{y}^{(a)T},  \ \ \ \text{for } a \in \{1,2,3,4\}.
\end{equation*}

To establish the validity of randomization-based inference based on $\widehat{\tau}_{a}$ and $S^2(Q)_a$, we consider the following regularity conditions, which are generalizations of some standard regularity conditions adopted in matched studies \citep{fogarty2018mitigating, zhang2024bridging, zhu2024bias,frazier2024bias} from the single treatment case to factorial treatments.

\begin{condition}\label{condition: bounded design matrix}
    (Bounded Entries of the Design Matrix $Q$): Let $q_{il}$ denote the entry in row $i$ and column $l$ of the design matrix $Q$ in $S^2(Q)_a$, for all $i=1,\cdots,I$ and $l=1,\cdots,L$. Then, we assume there exists a finite constant $C_1 \leq \infty$ such that $|q_{il}|\leq C_1$ for all $i=1,\dots,I, l=1,\dots,L$.
\end{condition}

\begin{condition}\label{condition: bounded fourth and no extreme}
    (Bounded Fourth Moments and No Extreme Matched Sets): For each matched set $i$, define $V_i^{(a)+}=\text{max}_{\mathbf{Z}_i\in\mathcal{Z}_i}V_i^{(a)}$, $V_i^{(a)-}=\text{min}_{\mathbf{Z}_i\in\mathcal{Z}_i}V_i^{(a)}$, and $M_i=V_i^{(a)+}-V_i^{(a)-}$ for $a=1,2,3,4$. We assume there exists a finite constant $C_2<\infty$ such that $I^{-1}\sum_{i=1}^{I}M_i^4\leq C_2$,$I^{-1}\sum_{i=1}^{I}q_{il}^4\leq C_2$, $I^{-1}\sum_{i=1}^{I}(V_i^{(a)+})\leq C_2$, and $I^{-1}\sum_{i=1}^{I}(V_i^{(a)-})\leq C_2$ for all $a=1,2,3,4$. Also, we have $\text{max}_{1\leq i \leq I}M_{i}^2/\{\sum^{I}_{i=1}M_{i}^2\} \to 0$ as $I \to \infty$.
\end{condition}

\begin{condition}\label{condition: convergence of means}
    (Convergence of Finite-Population Means): For each matched set $i$, let $\mu_i^{(a)}=E(V_i^{(a)}|\mathcal{Z})$ and $\nu_i^{(a
    )2}=\text{Var}(V_i^{(a)}|\mathcal{Z})$ for $a=1,2,3,4$. As $I\rightarrow\infty$, the following conditions hold for all $a=1,2,3,4$: (1) $I^{-1}\sum_{i=1}^{I}\mu_i^{(a)}$ and $I^{-1}\sum_{i=1}^{I}\mu_i^{(a)2}$ converge to some finite values; (2) $I^{-1}\sum_{i=1}^{I}\nu_i^{(a)2}$ converges to a finite positive value; (3) for each $l=1,\cdots,L$, the $I^{-1}\sum_{i=1}^{I}\mu_i^{(a)}q_{il}$ converges to a finite value; and (4) $I^{-1}Q^TQ$ converges to a finite, invertible $L\times L$ matrix $\widetilde{Q}$.
\end{condition}

Under the above regularity Conditions, Theorem~\ref{thm: CI validity} shows that the confidence intervals constructed from $\widehat{\tau}_a$ and $S^2(Q)_a$ are asymptotically valid for the proposed generalized factorial Neyman-type estimands $\tau_a$ ($a\in \{1,2,3,4\}$) without any modeling assumptions on super-populations or treatment effects.

\begin{theorem}\label{thm: CI validity}
  Conditional on matching and the ignorability assumption, as well as the regularity conditions stated in Conditions \ref{condition: bounded design matrix}--\ref{condition: convergence of means}, the confidence interval $[\widehat{\tau}_a-\Phi^{-1}(1-\alpha/2)S(Q)_a,\widehat{\tau}_a+\Phi^{-1}(1-\alpha/2)S(Q)_a]$ achieves an asymptotic coverage rate of at least $100(1-\alpha)\%$ for $\tau_a$, for $a \in \{1,2,3,4\}$ and any prespecified matrix $Q$. Here, $\Phi$ denotes the cumulative distribution function of the standard normal distribution, and $\alpha \in (0,0.5)$ is some prespecified level.
\end{theorem}

\section{Data Applications: Evaluating the Impacts of COVID-19 Social Distancing Practices in the United States}\label{sec: application}

During the COVID-19 pandemic, social distancing emerged as a central public health strategy, implemented through a broad array of policies and behavioral adaptations that altered multiple dimensions of human mobility \citep{Yeom2021social,Sun2022social}.  At the county level, these behaviors manifested in at least two distinct and substantively meaningful components--reductions in work trips and reductions in non-work trips--which respectively capture changes in professional versus recreational and social movement. Because these components reflect different pathways through which interpersonal contact is reduced, they may exert heterogeneous and/or interacting effects on health outcomes. Thus, any attempt to estimate the causal impact of ``social distancing'' must treat these mobility dimensions as a factorial treatment (e.g., reductions in work trips and reductions in non-work trips), rather than collapsing them into a single composite measure. However, existing matched observational studies (one of the most widely used observational study designs) lacked valid matching designs capable of handling multiple continuous treatments, leading prior work to rely on ad hoc weighted aggregations of mobility indicators \citep{zhang2023social, frazier2024bias}. As discussed in Section~\ref{sec: review}, such aggregation can distort the underlying treatment structure, obscure interactions, and render potential outcomes ill-defined. In this section, we apply our proposed universal factorial matching and inference framework to address these gaps in COVID-19 social distancing research and estimate the county-level causal effects of work- and non-work-trip reductions on COVID-19-related and drug-related outcomes in the United States.

To support this investigation, we integrate data from the University of Maryland's COVID-19 Impact Analysis Platform, the United States Census Bureau, the County Health Rankings and Roadmap Program, and the Centers for Disease Control and Prevention's (CDC's) county-level mortality data \citep{zhang2021platform,Remington2015ranking,acs5year,cdc_county_mortality}. We focus specifically on two distinct continuous county-level social distancing measures, defined as the average per-person percentage reductions in work and non-work trips during the pandemic period (February 16th, 2020, to April 20th, 2021), compared to the pre-pandemic baseline (February 1st to February 15th, 2020). These two metrics were selected as they represent the primary dimensions of mobility, which are professional versus recreational and social movement, and thereby capture the breadth of social distancing adherence. We examine several outcomes in this study. The primary outcome for this analysis is the cumulative number of confirmed COVID-19 cases during the pandemic (February 16th, 2020, to April 20th, 2021). Secondary outcomes include the cumulative number of COVID-19-related deaths and overdose-related deaths over the same period.

In the matching design, we implement the optimal two-stage non-bipartite matching method proposed in Section \ref{sec: matching} to construct strata of counties with similar covariates and four treatment patterns (i.e., four types of treatment component combination): $(h, h)$, $(h, l)$, $(l, h)$, and $(l, l)$. In the first stage, we match on the percentage reduction in work trips. When calculating the distance in the second stage, the percentage reduction in work trips is placed in the numerator, and the percentage reduction in non-work trips is placed in the denominator. This procedure results in 774 matched strata, each consisting of four counties. Specifically, we match on the following county-level baseline covariates: total population, population density, female (\%), under 18 (\%), over 65 (\%), median household income, unemployment (\%), some college (\%), Hispanic (\%), African American (\%), poverty rate (\%), smoking (\%), excess drinking (\%), flu vaccination (\%), number of membership associations per 10,000 people, traffic volume, driving alone to work (\%), and rural (0/1). We perform exact matching on ``rural (0/1)'' and balance all other covariates. Sinks are added to eliminate counties such that all the matched strata contained counties with the four treatment patterns as described in Table~\ref{table: patterns}. Following the guidance of \citet{rubin2007matching}, we perform matching without access to outcome data to avoid bias and maintain objectivity. 

After matching, the data consists of four distinct groups: $(h,h)$, $(h,l)$, $(l,h)$, and $(l,l)$. To assess covariate balance after matching, we calculate the standardized mean differences of all covariates for every pair of these four groups. To illustrate the performance of the matching procedure, we also create four corresponding pre-matching groups with the same $(h,h)$, $(h,l)$, $(l,h)$, and $(l,l)$ labels, defined according to the two continuous treatment doses in each group. Specifically, in pre-matching covariate balance evaluation, each continuous treatment dose is dichotomized at its median; in contrast, our framework does not need to dichotomize the treatment component. To avoid duplication, we present only the covariate balance before and after the optimal two-stage non-bipartite matching for the $(l,h)$ vs. $(l,l)$ groups; the covariate balance tables for all other contrasts (i.e., $(h, h)$ vs. $(h, l)$; $(h, h)$ vs. $(l,h)$; $(h, h)$ vs. $(l, l)$; $(h, l)$ vs. $(l, h)$; and $(h, l)$ vs. $(l, l)$) can be found in the supplementary materials. As shown in Table~\ref{table: balance}, after matching, the percentage reductions in work trips are very similar, while the percentage reductions in non-work trips differ substantially between the two groups. This suggests that our matching design can effectively produce the treatment pattern of $(l,h)$ vs. $(l,l)$ (as well as other treatment combination contrasts). Meanwhile, when examining covariate similarity, we find systematic differences between the $(l,h)$ and $(l,l)$ groups before matching, with around half of the absolute standardized mean differences exceeding 0.1. By contrast, after matching, all absolute standardized mean differences are less than 0.08, indicating excellent post-matching balance according to commonly used thresholds 0.2 or 0.1 \citep{rosenbaum2020design,zhu2024bias,zhang2023social}. The covariate balance results for the other group comparisons lead to similar conclusions and are provided in the Supplementary Materials.

\begin{table}[H]
    \caption{Pre-matching and post-matching covariate balance of $(l,h)$ VS. $(l,l)$. }
    \centering
    \resizebox{0.72\textwidth}{!}{
    \begin{tabular}[t]{ccccccc}
\toprule
  & \multicolumn{3}{c}{Pre-matching} & \multicolumn{3}{c}{Post-matching} \\
  \cline{2-7} \\ [-2ex]
 Covariate & \shortstack{$(l, h)$ \\ (n=765)} & \shortstack{$(l,l)$ \\ (n=785)} & Std.dif & \shortstack{$(l,h)$ \\ (n=774)} & \shortstack{$(l, l)$ \\ (n=774)} & Std.dif\\
\midrule
            \multicolumn{1}{l}{Total Population} & \multicolumn{1}{l}{$28985$} &
            \multicolumn{1}{l}{$22967$} &
            \multicolumn{1}{l}{$\;\;\:0.20$} &
            \multicolumn{1}{l}{$100874$} &
            \multicolumn{1}{l}{$88720$} &
            \multicolumn{1}{l}{$\;\;\:0.03$} \\
            \multicolumn{1}{l}{Pop Density} & \multicolumn{1}{l}{$65.1$} & \multicolumn{1}{l}{$51.3$} & \multicolumn{1}{l}{$\;\;\:0.06$} & 
            \multicolumn{1}{l}{$283$} & \multicolumn{1}{l}{$220$} & \multicolumn{1}{l}{$\;\;\:0.03$}\\
            \multicolumn{1}{l}{Female} & \multicolumn{1}{l}{$0.496$} & \multicolumn{1}{l}{$0.498$} & \multicolumn{1}{l}{$-0.07$} & 
            \multicolumn{1}{l}{$0.499$} & \multicolumn{1}{l}{$0.499$} & \multicolumn{1}{l}{$\;\;\:0.01$}\\
            \multicolumn{1}{l}{Under 18} & \multicolumn{1}{l}{$0.222$} & \multicolumn{1}{l}{$0.224$} & \multicolumn{1}{l}{$-0.05$} &
            \multicolumn{1}{l}{$0.223$} & \multicolumn{1}{l}{$0.222$} & \multicolumn{1}{l}{$\;\;\:0.02$} \\
            \multicolumn{1}{l}{Over 65} & \multicolumn{1}{l}{$0.198$} & \multicolumn{1}{l}{$0.200$} & \multicolumn{1}{l}{$-0.04$} &
            \multicolumn{1}{l}{$0.191$} & \multicolumn{1}{l}{$0.189$} & \multicolumn{1}{l}{$\;\;\:0.04$} \\
            \multicolumn{1}{l}{Median Household Income} &
            \multicolumn{1}{l}{$48548$} & \multicolumn{1}{l}{$50005$} & \multicolumn{1}{l}{$-0.14$} &
            \multicolumn{1}{l}{$53495$} & \multicolumn{1}{l}{$52703$} & \multicolumn{1}{l}{$\;\;\:0.06$} \\
            \multicolumn{1}{l}{Unemployment} & \multicolumn{1}{l}{$0.0571$} & \multicolumn{1}{l}{$0.0481$} & \multicolumn{1}{l}{$\;\;\:0.33$} & \multicolumn{1}{l}{$0.0527$} & \multicolumn{1}{l}{$0.0520$} & \multicolumn{1}{l}{$\;\;\:0.02$} \\  
            \multicolumn{1}{l}{Some College} & \multicolumn{1}{l}{$0.540$} & \multicolumn{1}{l}{$0.572$} & \multicolumn{1}{l}{$-0.28$} & \multicolumn{1}{l}{$0.580$} & \multicolumn{1}{l}{$0.582$} & \multicolumn{1}{l}{$-0.02$} \\
            \multicolumn{1}{l}{Hispanic} & \multicolumn{1}{l}{$0.120$} & \multicolumn{1}{l}{$0.065$} & \multicolumn{1}{l}{$\;\;\:0.39$} & \multicolumn{1}{l}{$0.097$} & \multicolumn{1}{l}{$0.089$} & \multicolumn{1}{l}{$\;\;\:0.06$} \\
            \multicolumn{1}{l}{African American} & \multicolumn{1}{l}{$0.108$} & \multicolumn{1}{l}{$0.051$} & \multicolumn{1}{l}{$\;\;\:0.41$} & \multicolumn{1}{l}{$0.089$} & \multicolumn{1}{l}{$0.087$} & \multicolumn{1}{l}{$\;\;\:0.02$} \\ 
            \multicolumn{1}{l}{Poverty} & \multicolumn{1}{l}{$0.162$} & \multicolumn{1}{l}{$0.144$} & \multicolumn{1}{l}{$\;\;\:0.31$} & \multicolumn{1}{l}{$0.146$} & \multicolumn{1}{l}{$0.145$} & \multicolumn{1}{l}{$\;\;\:0.01$}\\
            \multicolumn{1}{l}{Adult Smoking} & \multicolumn{1}{l}{$0.177$} & \multicolumn{1}{l}{$0.179$} & \multicolumn{1}{l}{$-0.04$} & \multicolumn{1}{l}{$0.173$} & \multicolumn{1}{l}{$0.176$} & \multicolumn{1}{l}{$-0.07$} \\
            \multicolumn{1}{l}{Excess Drinking} & \multicolumn{1}{l}{$0.165$} & \multicolumn{1}{l}{$0.174$} & \multicolumn{1}{l}{$-0.29$} & \multicolumn{1}{l}{$0.174$} & \multicolumn{1}{l}{$0.176$} & \multicolumn{1}{l}{$-0.06$} \\
            \multicolumn{1}{l}{Flu Vaccination} & \multicolumn{1}{l}{$0.386$} & \multicolumn{1}{l}{$0.401$} & \multicolumn{1}{l}{$-0.15$} & \multicolumn{1}{l}{$0.415$} & \multicolumn{1}{l}{$0.415$} & \multicolumn{1}{l}{$\;\;\:0.00$} \\
            \multicolumn{1}{l}{Social Association} & \multicolumn{1}{l}{$11.7$} & \multicolumn{1}{l}{$13.6$} & \multicolumn{1}{l}{$-0.30$} & \multicolumn{1}{l}{$11.8$} & \multicolumn{1}{l}{$11.8$} & \multicolumn{1}{l}{$\;\;\:0.01$} \\
            \multicolumn{1}{l}{Traffic Volume} & \multicolumn{1}{l}{$64.6$} & \multicolumn{1}{l}{$79.7$} & \multicolumn{1}{l}{$-0.13$} & \multicolumn{1}{l}{$130$} & \multicolumn{1}{l}{$131$} & \multicolumn{1}{l}{$-0.01$} \\
            \multicolumn{1}{l}{Drive Alone to Work} & \multicolumn{1}{l}{$0.797$} & \multicolumn{1}{l}{$0.797$} & \multicolumn{1}{l}{$\;\;\:0.00$} & \multicolumn{1}{l}{$0.797$} & \multicolumn{1}{l}{$0.802$} & \multicolumn{1}{l}{$-0.08$} \\
            \multicolumn{1}{l}{Non Metro} & \multicolumn{1}{l}{$0.796$} & \multicolumn{1}{l}{$0.829$} & \multicolumn{1}{l}{$-0.09$} & \multicolumn{1}{l}{$0.625$} & \multicolumn{1}{l}{$0.625$} & \multicolumn{1}{l}{$\;\;\:0.00$} \\
            \midrule
            \multicolumn{1}{l}{Work Trip} & \multicolumn{1}{l}{$9.00$} & \multicolumn{1}{l}{$8.93$} & \multicolumn{1}{l}{$\;\;\:0.00$} & \multicolumn{1}{l}{$13.9$} & \multicolumn{1}{l}{$15.5$} & \multicolumn{1}{l}{$-0.04$} \\
            \multicolumn{1}{l}{Non-work Trip} & \multicolumn{1}{l}{$1.122$} & \multicolumn{1}{l}{$-13.00$} & \multicolumn{1}{l}{$\;\;\:2.01$} & \multicolumn{1}{l}{$-1.59$} & \multicolumn{1}{l}{$-9.59$} & \multicolumn{1}{l}{$\;\;\:1.02$} \\
\bottomrule
\end{tabular}}
\label{table: balance}
\end{table}

Summarizing the results across all six tables, we find that there is little difference in the treatment variables between the two groups when they are both at high or low levels, as the corresponding absolute standardized mean differences are all below 0.1. For example, as shown in Table~\ref{table: similar}, in the comparison between $(h,h)$ and $(l,h)$ after matching, the absolute standardized mean difference for the percentage reduction in non-work trips (i.e., the contrast of the second treatment component: $(\cdot, h)$ vs. $(\cdot, h)$) is 0.05, indicating that this treatment dose is not meaningfully different between these two groups. Such a strong balance supports the validity and interpretability of our proposed generalized Neyman-type estimands. Meanwhile, in the post-matching comparison between $(h,h)$ and $(l,h)$, the absolute standardized mean difference for the percentage reduction in work trips (i.e., the contrast of the first treatment component: $(h, \cdot)$ vs. $(l, \cdot)$) is 0.77, indicating sufficient heterogeneity to detect the potential effects of reductions in work trips conditional on low levels of reductions in non-work trips. 

\begin{table}[ht]
\centering
    \caption{The treatment dose contrast (standardized difference in means) of reductions in work trips and reductions in non-work trips across the six treatment pattern comparisons.}
    \begin{tabular}{l|cc}
    \hline
    \textbf{} & Work trip & Non-work trip \\
    \hline
   $(h,h)$ vs. $(h,l)$ & -0.08 & 0.92 \\
   $(h,h)$ vs. $(l,h)$ & 0.77 & -0.05 \\
   $(h,h)$ vs. $(l,l)$ & 0.42 & 0.94 \\
    $(h,l)$ vs. $(l,h)$ & 0.82 & -0.99 \\
    $(l,h)$ vs. $(l,l)$ & -0.04 & 1.02 \\
    $(h,l)$ vs. $(l,l)$ & 0.45 & -0.05 \\
\hline
\end{tabular}
\label{table: similar}
\end{table}

After matching, for both COVID-19-related and drug-related outcomes, we will conduct a Fisher-type randomization test to test Fisher's sharp null hypothesis of no effect for all study units (counties). Then, we will calculate the point estimate and 95\% confidence interval for each of the four generalized factorial Neyman-type estimands (i.e., $\tau_{1}$, $\tau_{2}$, $\tau_{3}$, and $\tau_{4}$), two for main effects of the two treatment components (reductions in work trips and reductions in non-work trips) and two for their interaction effects.

\section{Discussion}\label{sec: discussion}

In this work, we propose a design and inference framework for factorial matched observational studies (i.e., observational studies involving multiple treatment variables), universally applicable to general treatment variables (binary, ordinal, continuous). The motivation for this study arises from the limitations of existing design-based approaches for observational studies with factorial treatments: when handling multiple non-binary (e.g., continuous) treatments, prior studies often either (i) dichotomize continuous treatments to facilitate the classic binary-treatment factorial matching designs or (ii) combine multiple treatments into a single composite variable and employ the classic single-treatment matching designs, both of which can introduce bias and hinder valid and interpretable estimation and inference. To clearly convey the key ideas of our framework without notational complexity, we illustrate the design and inference procedures using a bivariate treatment case, focusing on the universal matching design, the generalized factorial Neyman-type estimands, and the subsequent randomization-based inference. Nevertheless, the proposed framework can be generalized to settings with more than two treatment variables. An example with three treatments is provided in the Appendix. To our knowledge, this is the first model-free and design-based (finite-population-based) framework for observational studies involving multiple, possibly non-binary treatments. At this stage, this manuscript serves as a methodological protocol. We have not yet accessed or analyzed the outcome data, and therefore, no empirical results are presented here. The current version focuses on the conceptual and methodological development of the framework and will be updated once the data analysis is completed.

\section*{Acknowledgments}

This work is in part supported by NIH Grant R21DA060433.

\bibliographystyle{apalike}
\bibliography{references}

\end{document}